\documentclass[a4paper,12pt]{article}

\usepackage{graphicx}
\usepackage{amsmath,amssymb}
\usepackage{a4wide}
\usepackage{cite}
\usepackage{subfigure}
\usepackage{color}

\renewcommand{\baselinestretch}{1.5}

\begin{document}

\begin{center}
\renewcommand{\baselinestretch}{1.3}\small\normalsize

   {\large \bf Dissociative adsorption of methane on \\
               surface oxide structures of Pd-Pt alloys}

   \vspace{0.5cm}
    Arezoo Dianat,$^{1,*}$
    Nicola Seriani,$^{2}$
    Lucio Colombi Ciacchi,$^{3,4}$ \\
    Wolfgang Pompe,$^{1}$
    Gianaurelio Cuniberti,$^{1}$
    Manfred Bobeth$^{1}$

   \vspace{0.25cm}
   {\it
       $^1$Institute for Materials Science and Max Bergmann Center of Biomaterials 
          , Dresden University of Technology, D-01062 Dresden, Germany \\
       $^2$Fakult\"at Physik, Universit\"at Wien,
           Sensengasse 8, 1090 Wien, Austria \\
       $^3$Hybrid Materials Interfaces Group, Faculty of Production Engineering \\
          and Bremen Centre for Computational Materials Science, \\
          University of Bremen, D-28359 Bremen, Germany\\
      $^4$Fraunhofer Institute for Manufacturing Technology and Applied Material Research IFAM, \\
          D-28359 Bremen, Germany\\
}   
\vspace{0.5cm} \today

\end{center}

\begin{abstract}

The dissociative adsorption of methane on variously oxidized Pd, Pt 
and Pd-Pt surfaces is investigated using density-functional theory, as a 
step towards understanding the combustion of methane on these materials.
For Pd-Pt alloys, models of surface oxide structures are built on the basis 
of known oxides on Pd and Pt.  
The methane adsorption energy presents large variations depending on the
oxide structure and composition.
Adsorption is endothermic on the bare Pd(111) metal surface as well as 
on stable thin layer oxide structures such as the ($\sqrt{5}\times\sqrt{5}$) 
surface oxide on Pd(100) and the PtO$_2$-like oxide on Pt(111).
Instead, large adsorption energies are obtained for the (100) surface of bulk PdO,
for metastable mixed Pd$_{1-x}$Pt$_x$O$_{4/3}$ oxide layers on Pt(100), 
and for Pd-Pt(111) surfaces covered with one oxygen monolayer.
In the latter case, we find a net thermodynamic preference for a
direct conversion of methane to methanol, which remains adsorbed 
on the oxidized metal substrates via weak hydrogen-bond interactions.
\end{abstract}

\vfill \noindent
$^*$ To whom correspondence should be addressed: arezoo.dianat@nano.tu-dresden.de

\newpage

\section{Introduction}

The catalytic oxidation of hydrocarbons in general, and of methane in particular, is 
considered as an effective method of power generation associated with low emissions 
of CO$_2$ and NO$_x$. 
Because of the high H/C ratio of methane, the heat of combustion per mole of 
generated CO$_2$ is higher than for other fuels, e.g. twice as much as for coal~\cite{chin}.  
The interest towards the development and optimization of novel catalysts for the 
combustion of methane has thus considerably increased over the last years. 
Especially palladium-based catalysts have been extensively explored because 
of their high catalytic oxidation activity. 
Among those catalysts, the bimetallic system Pd-Pt has been the object of  
many investigations~\cite{Micheaud,Narui,Ersson,Deng,Yammamoto,Ozawa,Persson1,Persson2,Persson3,Lapisardi}.
In several studies it was found that Pd-Pt catalysts exhibit higher methane 
conversion efficiency~\cite{Narui,Lim,Ersson,Deng,Yammamoto,Persson1,Persson2,Persson3,Lapisardi} 
and better long-term stability~\cite{Narui,Ozawa,Lapisardi} than pure Pd.

The performance of Pd-Pt catalysts strongly depends on the chemical state of the metal
surface at the conditions suitable for oxidation, which in many cases lead to
the formation of superficial oxide phases.
For pure Pd and Pt catalysts, there is agreement that oxide formation 
has a positive effect on their methane oxidation activity~\cite{Kondo,Han}, although the 
active oxide phases have not been unequivocally determined yet.
On the one hand, some authors have proposed that bulk PdO is less active than either 
a thin oxide layer~\cite{Oh} or a layer of adsorbed oxygen on Pd~\cite{Hicks1,Hicks2}.
On the other hand, the formation of bulk oxide has been suggested 
to explain the observed increase in catalytic activity~\cite{Carlsson,Burch1}.  
Recently, Gabasch et al.~\cite{Gabasch1} reported that bulk PdO seeds 
grown on a surface, otherwise covered by a Pd$_5$O$_4$ surface oxide, are the 
active phase for methane oxidation on a Pd(111) single crystal surface.
The catalytic behavior of pure Pt metal is quite different.
According to investigations in Refs. \cite{Carlsson,Burch1,Becker,Zhdanov}, its catalytic 
activity reaches a maximum at a submonolayer coverage of adsorbed oxygen, and further 
oxidation with formation of PtO$_2$ leads to activity loss.
Still, it is unclear whether a similar behavior can be expected also for other
Pt oxide phases that could develop in an oxidizing atmosphere.
Platinum bulk oxide phases comprise $\alpha$-PtO$_2$, $\beta$-PtO$_2$, 
Pt$_3$O$_4$ and PtO ~\cite{Muller,Nicola2,Jacob}.
Correspondingly, a large variety of surface oxides can be expected~\cite{Nicola,Nicola3,Devarajan}. 
Previous calculations have suggested that Pt$_3$O$_4$ might be an active
phase for the catalytic oxidation of carbon compounds~\cite{Nicola}. 
Similarly, in the case of palladium, besides PdO also other bulk oxide phases 
could form, as e.g. PdO$_2$~\cite{Nolte}.
Different surface oxides have been observed on Pd(111)~\cite{Klikovits,Kan,Hinojosa}, 
Pd(100)~\cite{Kostelnik} and on stepped surfaces~\cite{Westerstrom}.

Compared with the large amount of data available for pure Pd and Pt, little is known 
about the oxidation behavior of the bimetal Pd-Pt and its 
relation to the higher catalytic activity displayed by these alloys. 
Persson et al.~\cite{Persson3} suggested that Pd-Pt catalysts on alumina consist 
of a PdO-rich phase coexisting with a palladium-rich Pd-Pt alloy.
Studying the composition dependence of the catalytic activity of Pd-Pt,
Lapisardi et al.~\cite{Lapisardi} found the highest activity for catalysts with
very high Pd content Pd$_{0.93}$Pt$_{0.07}$/Al$_2$O$_3$.

In a previous study~\cite{Dianat2}, we have investigated by means of density-functional
theory (DFT) the thermodynamic stability of mixed bulk oxides Pd$_{1-x}$Pt$_x$O$_y$ 
isostructural to known bulk oxide phases of pure Pd and Pt.
According to these calculations, the mixed oxide phase Pd$_{1-x}$Pt$_x$O$_2$
with crystal structure analogous to $\alpha$-PtO$_2$ is stable only
at rather low temperature ($<$ 500~K for atmospheric oxygen pressure).
With increasing temperature, a phase mixture of PdO with first PtO$_2$ 
and later Pt$_3$O$_4$ becomes stable.
At higher temperature, the oxides decompose directly to metallic Pd-Pt,
except for Pd-rich systems, where PdO and metallic Pt coexist within a small
temperature window.

In the present DFT study, in a first step we analyze the stability of oxidized  
Pd-Pt surfaces as a basis for investigating their interaction with methane.
Because of the lack of relevant experimental information, known surface oxide 
structures of pure Pd and pure Pt are 
used as guidelines to construct Pd-Pt surface oxide models.
In a second step, we compute the driving forces for the dissociative adsorption 
of CH$_4$ on various oxidized Pd-Pt surfaces, which is commonly believed to be 
the rate-determining reaction in the catalytic combustion of methane.

The paper is outlined as follows.
Computational details of the DFT analysis are briefly described in Sect. 2.
In Sect. 3, we report on the calculated oxygen binding energies of various 
Pd-Pt oxide structures.  
The calculated methane adsorption energies on these oxide 
structures are then presented in Sect. 4. 
Finally, our results are discussed and interpreted in relation to
experimental findings in Sect. 5.

\section{Computational details}

Our DFT calculations are performed by means of the Vienna ab initio simulation 
package (VASP)~\cite{Kresse3,Kresse1,Kresse4}, using the PBE generalized gradient 
approximation (GGA) for the exchange-correlation functional~\cite{PBE} and the 
PAW method~\cite{Blochl,Kresse2}.
The wave functions are expanded in plane waves up to a kinetic energy cut-off of 400 eV.
The periodically repeated simulation cells include slabs of six substrate layers covered 
with either adsorbed oxygen or a thin oxide layer, and with adsorbed CH$_3$ and H. 
In all simulations, the vacuum gap between the slab surface models is larger than 15~\AA.  
Unless stated otherwise, the size of the simulation cell corresponds to a 
(2$\times$2) surface unit cell of the metal substrate.
Integration in the first Brillouin zone is performed using Monkhorst-Pack
grids~\cite{Monkhorst} including 25 $k$-points in the irreducible wedge.
In all calculations, the positions of all atoms are optimized until all force components 
become less than $0.01$~eV/\AA. 
Convergence of energy differences with respect to the used cut-off energies and 
$k$-point grids is ensured within a tolerance of 10~meV/atom. 
Further computational details can be found in Ref.~\cite{Dianat1}.

\section{Surface oxide structures on Pd-Pt}

The oxidation of Pd and Pt surfaces proceeds from the chemisorption of oxygen atoms 
through the formation of surface oxides to the development of bulk oxide.
In the case of palladium, the structure of surface oxide phases has been the subject 
of many experimental~\cite{Ketteler,Todor2,Zemlyanov,Gabasch} as well as theoretical 
\cite{Reuter1,Todor2,Stampfl,Rogal, Klikovits, Kostelnik, Westerstrom} investigations.
Phase diagrams of surface oxide structures in dependence on the chemical potential of oxygen 
have been thoroughly characterized for several surface orientations~\cite{Klikovits, Kostelnik}. 
On Pd(111), chemisorbed oxygen at low coverage is arranged with a p(2$\times$2) periodicity.
At higher coverage, several surface oxide phases form and coexist, as observed
in STM investigations accompanied by theoretical modelling~\cite{Klikovits, Ketteler,Reuter1}. 
On the contrary, on Pd(100) only one surface oxide has been found~\cite{Todor2, Kostelnik}, consisting of two unit cells of PdO(101) over a ($\sqrt{5} \times\sqrt{5}$) Pd(100) cell.
Investigations of oxide formation on platinum revealed the following.  
On Pt(111), the formation of a bulk-like, strongly distorted $\alpha$-PtO$_2$ 
surface oxide was observed at an oxygen partial pressure of 0.5 atm and 
temperatures from 520-910~K~\cite{Ellinger, Weaver}.
$\alpha$-PtO$_2$ was predicted to be the stable low temperature phase also 
by DFT calculations~\cite{Nicola,Dianat2}.
On Pt(100), DFT calculations suggest the formation of an $\alpha$-PtO$_2$-like surface oxide at low temperature~\cite{Nicola, Nicola3} and the existence of a stability region for a Pt$_3$O$_4$-like oxide layer at higher temperature.

\subsection{On-surface and sub-surface adsorbed oxygen}

To characterize the binding strength of oxygen atoms adsorbed at metal surfaces,
we calculate an average oxygen binding energy $E_{b}$ per 
O atom according to the formula
\begin{equation}
	E_{b} = \frac{1}{N_{O}} \left[E_{O@S} - E_S - \frac{N_{O}}{2} E_{O_2}\right] \;,
\label{eq2}
\end{equation}
where $E_{O@S}$ is the total energy of the oxygen-metal system, 
$E_S$ the energy of the bare metal substrate, $E_{O_2}$ the energy 
of a free oxygen molecule, and N$_{O}$ the number of O atoms.
The oxygen atoms in formula (\ref{eq2}) include oxygen located on the surface as well as
in sub-surface positions.
The calculated oxygen binding energies on (111) surfaces are listed in Table~\ref{on_sub_tab}
for different oxygen coverages up to 1 monolayer (ML).
The values in the last column are obtained for a stack of atomic layers with 
Pd in the surface layer, Pt in the sub-surface layer, and Pd in the remaining layers. 
In a previous DFT study~\cite{Dianat2}, we have calculated oxygen adsorption 
energies on mixed Pd-Pt surface and sub-surface layers. 
For the considered Pd-Pt compositions and configurations, the oxygen adsorption 
energies have been found to vary between -1.26~eV and -0.94~eV. 
In the case of stacks of pure Pd and Pt layers, we have obtained 
an adsorption energy of -0.99~eV on Pt/Pd/Pd(111) and -1.26~eV on Pd/Pt/Pd(111).
The latter value presents the strongest oxygen binding that we have found on
all considered stacks of mixed metal layers.
This strong binding has been attributed to charge transfer from Pd to Pt which causes 
a strong binding between the more electropositive Pd atoms and electronegative 
O atoms.

The most stable oxygen adsorption sites on the (111) surface of Pd and Pt 
are the fcc hollow sites~\cite{Todor1,Getman}. 
According to the energy values in Table 1, the oxygen binding energies on 
pure Pt(111) are significantly smaller than on pure Pd(111) for all considered 
coverages.
The strongest oxygen binding is found for the Pd/Pt/Pd(111) layer stack.   

Further calculations are performed for the case of fixed sub-surface 
oxygen coverage of 0.25~ML, and increasing on-surface coverage up to 
0.75~ML, thus giving a maximum total coverage of 1~ML.
Between the first and second metal layer there are three high-symmetry 
positions: an octahedral site underneath the fcc on-surface hollow site, 
a tetrahedral site (tetra I) below the hcp on-surface hollow site, and a 
second tetrahedral site (tetra II) directly below a surface metal atom~\cite{Todor1}.
For simultaneous on-surface and sub-surface adsorption, the fcc hollow site on
the surface and the tetra I sub-surface position are found to be the most stable 
oxygen positions for all considered oxygen coverages.
The corresponding average oxygen binding energies are given in Table~\ref{on_sub_tab} 
(referred to as on+sub).
They reveal that, in addition to on-surface adsorption, oxygen incorporation underneath
the surface layer becomes favorable for a total oxygen coverage  
$\theta_{\rm tot} \ge$ 0.75~ML on Pd(111), and for $\theta_{\rm tot}$ = 1~ML on Pt(111). 
This result is in agreement with other DFT calculations for Pd(111)~\cite{Todor1}.
For solely on-surface adsorption, the binding of oxygen is slightly stronger
on Pd/Pt/Pd(111) than on Pd(111), whereas with sub-surface oxygen the average binding
energy is larger for pure Pd(111) due to the unfavorable binding of oxygen to the 
Pt sub-surface layer in the case of Pd/Pt/Pd(111).

\subsection{Thin oxide layers}

\begin{figure}[ht!]  
\begin{center}
\includegraphics[width=\textwidth]{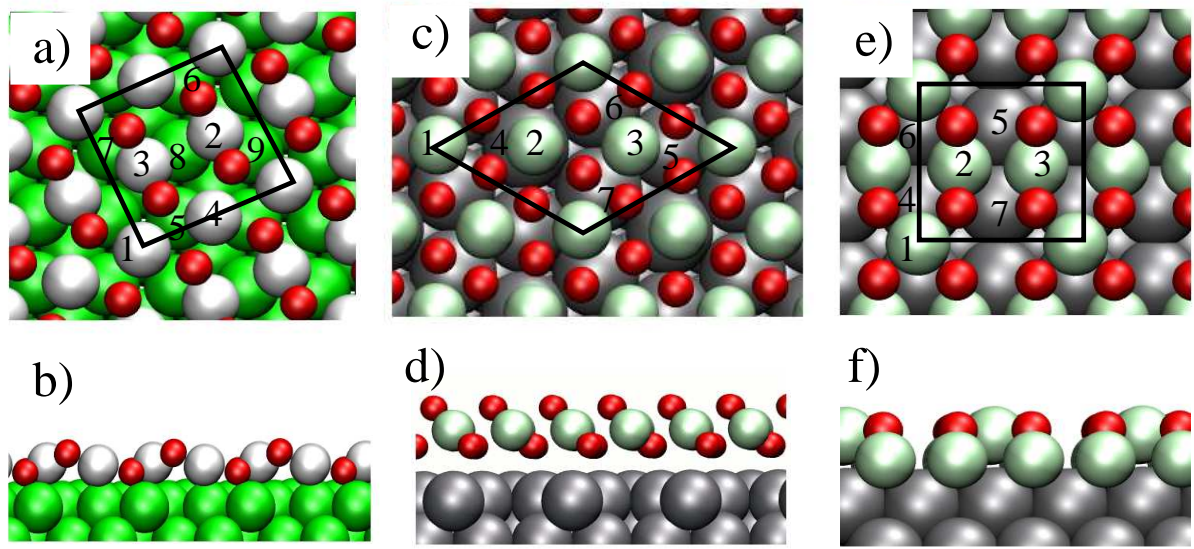}
\end{center}
\caption{\label{Thin_Film_Str}
Top- and side-views of optimized atomic structures of different oxide layers:
(a,b) PdO(101) on Pd(100), 
(c,d) $\alpha$-PtO$_2$(0001) on Pt(111), and (e,f) Pt$_3$O$_4$(100) on Pt(100) 
(O - red small spheres, metal - large spheres). 
In the case of mixed oxide and surface metal layers, Pd atoms are replaced by Pt  
(and vice versa) as described in the text.
}
\end{figure}

With increasing oxygen coverage, surface oxide phases start forming on
Pd and Pt surfaces. 
To our knowledge, there is no experimental information 
concerning the structure of such phases on the Pd-Pt bimetal.
For this reason, model structures of possible Pd-Pt surface oxides are 
constructed here on the basis of the known oxide structures of pure Pd and Pt.
In particular, we consider the experimentally observed PdO(101)-like~%
\cite{Todor2, Kostelnik} and the theoretically predicted Pt$_3$O$_4$-like~%
\cite{Nicola} oxide layers on the (100) surface, as well as the 
$\alpha$-PtO$_2$-like~\cite{Ellinger, Nicola} layer on the (111) surface.
According to our previous calculations~\cite{Dianat2}, these layers are 
characterized by small lattice misfits between oxide and  metallic substrate, 
namely 2\% for PdO(101)/Pd(100), 3.5\% for $\alpha$-PtO$_2$(0001)/Pt(111), 
and less than 1\% for Pt$_3$O$_4$(100)/Pt(100). 
The corresponding oxide structures obtained after DFT geometry optimization 
are shown in Fig.~\ref{Thin_Film_Str}.

In the case of PdO(101)/Pd(100), the Pd atoms of the oxide are located above 
the surface hollow sites of the substrate.
The oxygen amount in the oxide layer corresponds to a coverage of 0.8~ML, 
referred to the surface metal atoms.
In the case of $\alpha$-PtO$_2$(0001)/Pt(111), the Pt atoms of the oxide are located 
above top sites, fcc hollow sites, as well as hcp hollow sites of the substrate, 
and the oxygen coverage is 1.5~ML. 
The epitaxial Pt$_3$O$_4$(100)/Pt(100) oxide layer corresponds to an oxygen 
coverage of 1~ML.
In this case, the oxide layer includes two metal layers 
(cf. Figs. \ref{Thin_Film_Str}(e) and (f)).  

In the case of mixed Pd-Pt systems, we consider special Pd-Pt configurations 
in the oxide layers and in the metal surface layers which are generated by the 
following replacements. 
For example, starting from a pure Pd system in the case 
of Fig. 1a, we first replace Pd atom 1 in the oxide layer by Pt.
With increasing Pt concentration in this layer Pd atoms 2, 3 and 4 are
replaced by Pt.
The Pt concentration in the metal surface layer is increased in the same
manner starting with Pd atom 5. 
For the other two layer systems (Figs. 1c and 1e), the Pt atoms are replaced by 
Pd in analogous manner.

In order to characterize the stability of oxide structures with different Pd-Pt 
configurations and variable oxygen coverage, an average oxygen binding energy can
be defined as
\begin{equation}
	E_{b} = \frac{1}{N_O} \left[E_{Layer@S} - E_S - N_{Pd}^{Layer}E_{Pd}^{bulk} - 
				N_{Pt}^{Layer}E_{Pt}^{bulk} - \frac{N_O}{2} E_{O_2}\right] \;,
\label{Bin_sur_Ox}
\end{equation}
where $E_{Layer@S}$ is the total energy of the substrate together with the oxide layer, 
$E_S$ the energy of the substrate alone, $E_{Pd}^{bulk}$ and $E_{Pt}^{bulk}$ the energies 
of Pd and Pt atoms in their bulk phases, and $N_{O}$, $N_{Pd}^{Layer}$ and $N_{Pt}^{Layer}$ 
the numbers of oxygen, Pd, and Pt atoms in the oxide layer. 
The binding energy (\ref{Bin_sur_Ox}) corresponds to the energy gained by deposition 
of molecular oxygen and formation of an oxide with metal atoms from ideal bulk metal reservoirs.
With this definition we neglect the difference of chemical potentials of the
metal atoms in the pure metal bulk and in a Pd-Pt alloy. 
Since the free energy of mixing of Pd-Pt is of the order of few tens of meV per
metal atom, while the oxygen binding energies are of the order of one eV, 
this approximation will not significantly influence our conclusions.

\begin{figure}[ht!]
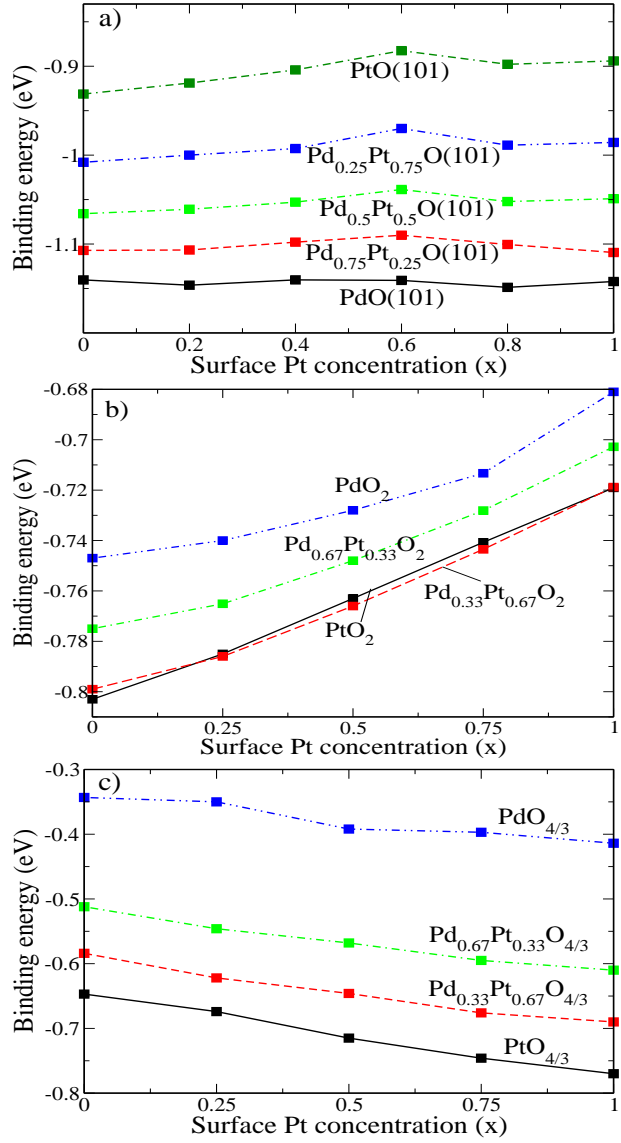
  
\begin{center}
\includegraphics[clip,width=8cm,height=5cm]{Fig2a.eps}\\
\includegraphics[clip,width=8cm,height=5cm]{Fig2b.eps}\\
\includegraphics[clip,width=8cm,height=5cm]{Fig2c.eps}
\end{center}
\caption{
Calculated average binding energies of oxygen for different oxide layer structures
as a function of the Pt concentration $x$ of the outermost substrate layer.
The different curves correspond to different compositions of the oxide layer,
as indicated by the labels.
(a) Pd$_{1-y}$Pt$_{y}$O(101) on Pd$_{1-x}$Pt$_{x}$/Pd(100), 
(b) $\alpha$-Pd$_{1-y}$Pt$_{y}$O$_2$(0001) on Pd$_{1-x}$Pt$_{x}$/Pt(111), and 
(c) Pd$_{1-y}$Pt$_{y}$O$_{4/3}$(100) on Pd$_{1-x}$Pt$_{x}$/Pt(100).
}
\label{Binding_Oxide}
\end{figure}

To study the effect of the Pd-Pt composition on the oxygen binding energy, 
we vary the Pd/Pt ratio both in the oxide layer and in the outermost 
substrate-layer.
All other substrate layers are composed of Pd in the case of 
the PdO(101) layer on Pd(100), and of Pt for the other two cases.
The calculated oxygen binding energies for the three oxide layers are shown 
in Fig.~\ref{Binding_Oxide} as a function of the Pt concentration of the 
outermost substrate layer. 
The different curves presented correspond to different compositions 
of the oxide layer.
Every curve point represents one special Pd-Pt configuration as 
noted above. 
Test calculations for another configuration at fixed composition
for Pd$_{0.5}$Pt$_{0.5}$O on Pd$_{0.4}$Pt$_{0.6}$/Pd(100) and
for $\alpha$-Pd$_{0.33}$Pt$_{0.67}$O$_2$ on Pd$_{0.75}$Pt$_{0.25}$/Pt(111)
yield only small differences in the average oxygen binding energy of 
0.012~eV and 0.006~eV, respectively.

For the PdO(101) layer structure (Fig.~\ref{Binding_Oxide}(a)), the maximum
variation of the oxygen binding energy with the substrate-layer composition
is less than 0.05~eV.
Similarly, the binding energy varies less than 0.1~eV for the $\alpha$-PtO$_2$ 
structure (Fig.~\ref{Binding_Oxide}(b), note the different energy scale). 
Replacement of Pd atoms of the PdO-like layer with Pt is energetically unfavorable.
Namely, the O binding energy changes from -1.15 to -0.90~eV passing from 
a pure PdO to a pure PtO layer.
The same holds for replacing Pt atoms of the $\alpha$-PtO$_2$ layer with Pd.
An analogous result has previously been found for mixed {\it bulk} oxides of 
Pd-Pt~\cite{Dianat2}.
For the case of the Pt$_3$O$_{4}$-like oxide layer (Fig.~\ref{Binding_Oxide}(c)), 
the changes in the binding energy with the substrate-layer composition are also 
small (less than 0.12 eV).
For this oxide layer it is very unfavorable to replace Pt atoms with Pd. 
Indeed, a Bader analysis~\cite{Bader, Henkelman} reveals that the oxygen atoms in 
Pd$_3$O$_{4}$/Pt(100) gain 0.67~e from metal atoms, while the corresponding value 
for Pt$_3$O$_{4}$/Pt(100) is 0.73~e, reflecting the stronger oxygen binding in 
the latter case. 
For the other two oxide structures considered, the charge transfer from metal atoms 
to oxygen does not change significantly with variation of the Pd-Pt composition in 
the oxide layer, consistently with the calculated small differences in the 
oxygen binding energy.

\section{Dissociative adsorption of methane}

The first important step in the catalytic oxidation of methane is its adsorption 
on the catalyst surface and dissociation into adsorbed methyl and hydrogen.  
To get a first insight into the catalytic activity of Pd-Pt catalysts for methane 
oxidation, we compute the thermodynamic driving force for this 
adsorption reaction on various oxide structures, in particular on the
superficial oxides of Pd-Pt considered in the previous section.   
The adsorption energy for the dissociative adsorption of CH$_4$ to CH$_3$ and H is defined as
\begin{equation}
	E_{a} =E_{{CH_3}\&H@S} - E_S - E_{CH_{4}} \;,
\label{equ_diss}
\end{equation}
where $E_{{CH_3}\&H@S}$ is the total energy of methyl and hydrogen adsorbed on the substrate,
$E_S$ the energy of the substrate, and $E_{CH_{4}}$ the energy of methane in the gas phase.
We do not perform zero point energy corrections of the calculated 
adsorption energies since we think 
that these corrections lead to comparatively small shifts of energy values
(see e.g. Ref.~\cite{Novell}), which is of minor importance for 
comparing adsorption energies on different systems as main goal
of this study.

\subsection{CH$_4$ adsorption on clean metal surfaces}

We first report on our calculations concerning methane adsorption on 
pure metallic (111) and (100) surfaces. 
In agreement with other DFT calculations~\cite{Au1,Paul}, we find 
that on the (111) surface the most stable adsorption site for CH$_3$ is 
on top of metal atoms and for H on fcc hollow sites. 
This applies to pure Pd(111) and Pt(111) as well as to the Pd/Pt/Pd(111) layer stack 
with Pd in the surface and Pt in the sub-surface atomic layer.
The calculated adsorption energies, shown in Table~\ref{diss}, are -0.08~eV for 
the Pt(111) surface and 0.19~eV for the Pd(111) surface. 
The positive value for Pd(111) corresponds to an endothermic reaction. 
This is in agreement with previous DFT calculations for methane adsorption on Pd(111) within the generalized gradient approximation (GGA) ($E_{a}$ = 0.27 eV), whereas an
exothermic reaction was found by using the local density approximation (LDA) 
($E_{a}$ = -0.62 eV)~\cite{Paul}. 
For comparison, at the LDA level we obtain adsorption energies of 
-0.43~eV, -0.66~eV, and -0.45~eV on Pd(111), Pt(111), and Pd/Pt/Pd(111), respectively.
Thus, for all considered systems, the LDA values are shifted roughly by -0.6~eV 
with respect to the GGA values.

In the case of all (100) surfaces, CH$_3$ on top of metal atoms and H on bridge sites are the most stable adsorption configurations.
On Pd(100) and Pd/Pt/Pd(100), dissociative adsorption of methane is found 
to be endothermic, in agreement with DFT calculations in~\cite{Zhang}. 
Only on Pt(100), methane adsorption is exothermic with an adsorption 
energy of -0.31 eV.

\subsection{CH$_4$ adsorption on surfaces with adsorbed oxygen}

\begin{figure}[ht!]  
\begin{center}
\includegraphics[clip,width=9.5cm,height=9.5cm]{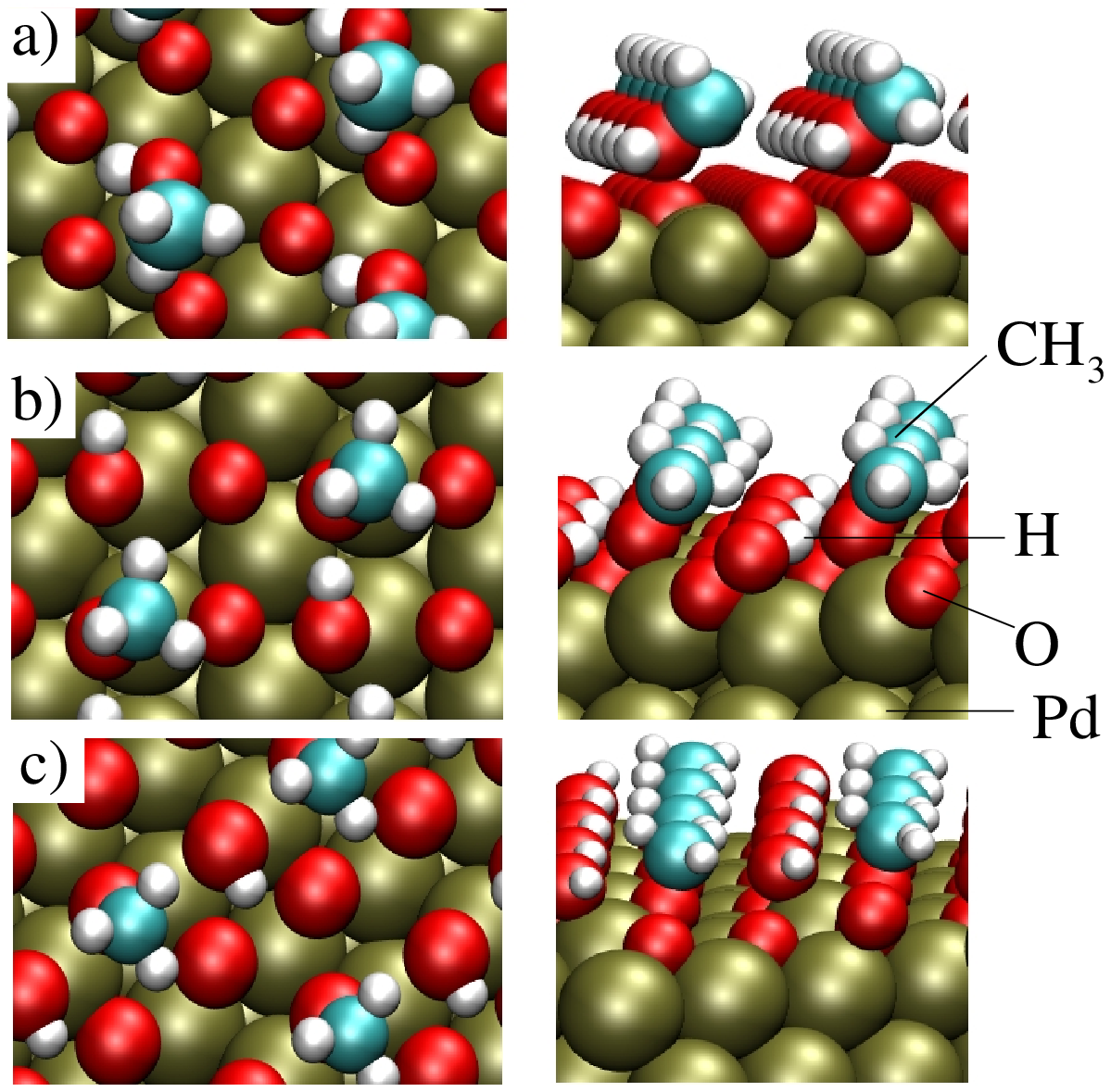}
\end{center}
\caption{
Optimized atomic structures of adsorbed CH$_3$ and H on a Pd(111) metal surface 
covered with 1 ML oxygen.
In the calculations the initial lateral positions of CH$_3$ and H have been chosen
as follows: a) CH$_3$ on top of a metal atom and H on top of oxygen, 
b) CH$_3$ and H on top of oxygen atoms, and c) CH$_3$ on hcp hollow site and 
H on top of oxygen.  
The largest adsorption energy is obtained for a). 
Distances between adsorbed atoms and nearest neighbor substrate atoms in~\AA:  
a) d$_{C-Pd}$ = 4.00, d$_{C-O}$ = 1.43, d$_{H-O}$ = 0.98; 
b) d$_{C-Pd}$ = 3.04, d$_{C-O}$ = 1.41, d$_{H-O}$ = 1.00; 
c) d$_{C-Pd}$ = 3.05, d$_{C-O}$ = 1.40, d$_{H-O}$ = 0.98.
}
\label{CH4_config_Pd}
\end{figure}

\begin{figure}[ht!]  
\begin{center}
\includegraphics[clip,width=9.5cm,height=9.5cm]{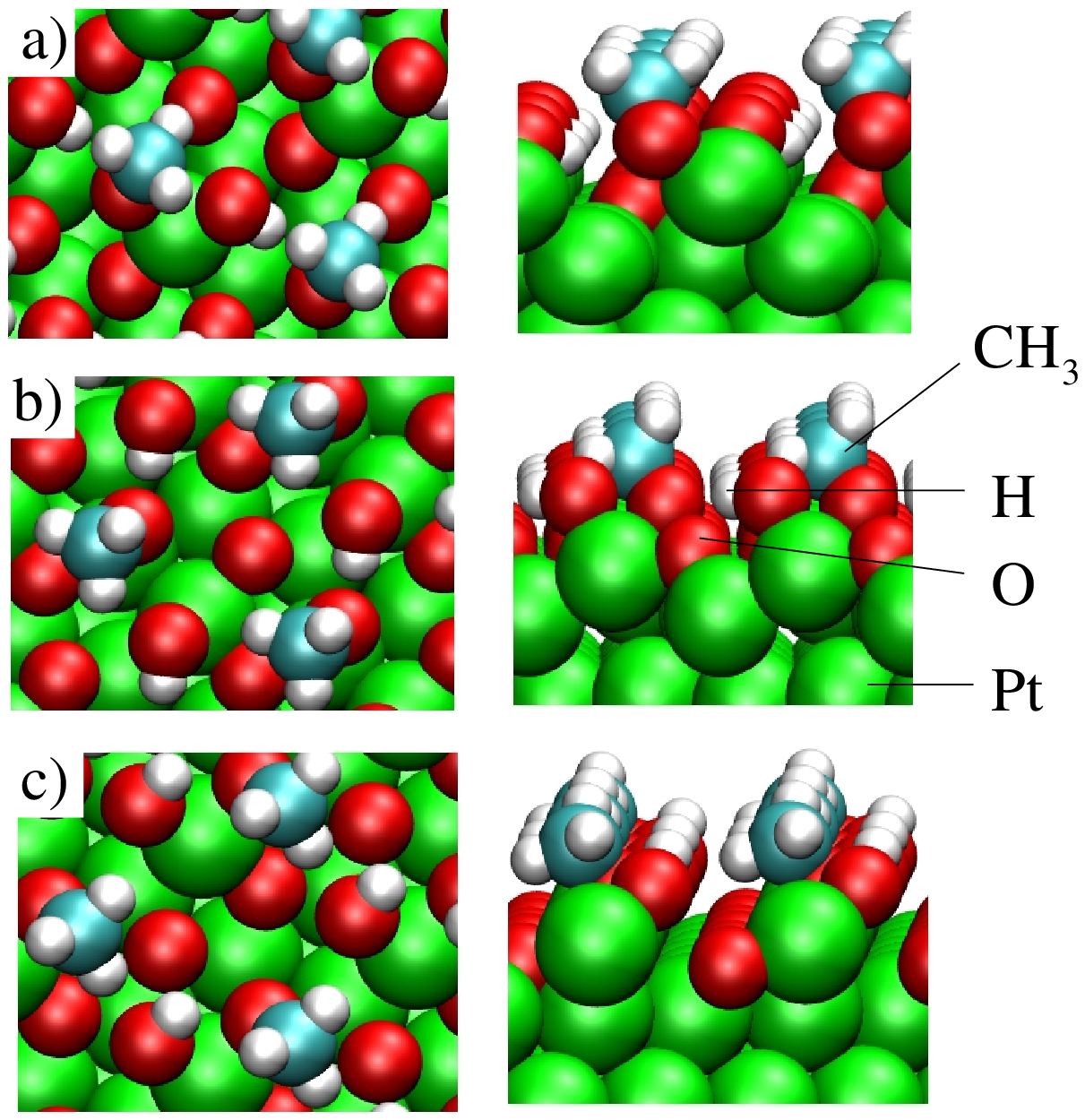}
\end{center}
\caption{
Optimized atomic structures of adsorbed CH$_3$ and H on a Pt(111) metal surface 
covered with 1 ML oxygen.
In the calculations the initial lateral positions of CH$_3$ and H have been chosen
as follows: a) CH$_3$ on top of a metal atom and H on top of oxygen, 
b) CH$_3$ and H on top of oxygen atoms, and c) CH$_3$ on hcp hollow site 
and H on top of oxygen.  
The largest adsorption energy is obtained for a). 
Distances between adsorbed atoms and nearest neighbor substrate atoms in~\AA:
a) d$_{C-Pd}$ = 2.97, d$_{C-O}$ = 1.42, d$_{H-O}$ = 0.99; 
b) d$_{C-Pd}$ = 3.03, d$_{C-O}$ = 1.41, d$_{H-O}$ = 1.00; 
c) d$_{C-Pd}$ = 2.92, d$_{C-O}$ = 1.41, d$_{H-O}$ = 1.00.
}
\label{CH4_config_Pt}
\end{figure}

After analyzing methane adsorption on clean metal surfaces, we investigate here
the adsorption on metal surfaces covered with 1 ML oxygen. 
A first set of calculations is performed starting with 1~ML of oxygen
adsorbed solely {\it on} the (111) surface.
As initial condition for the DFT calculations, the following positions 
of adsorbed CH$_3$ and H are chosen:
H always on top of oxygen atom, and 
(i) CH$_3$ on top of metal atom,
(ii) CH$_3$ on top of oxygen atom, and
(iii) CH$_3$ on hcp hollow site.
For these three cases, the calculated adsorption energies after 
structural relaxation are listed in Table~\ref{diss}. 
The corresponding optimized atomic structures are shown in Figs.~\ref{CH4_config_Pd}
and~\ref{CH4_config_Pt} for the Pd(111) and Pt(111) surfaces.
The structures obtained for the Pd/Pt/Pd(111) layer stack are qualitatively 
the same as for Pd(111).
The largest adsorption energies are obtained starting with CH$_3$ on 
top of a metal atom and H on top of an oxygen atom.
In particular, the adsorption energy on Pd/Pt/Pd(111) is slightly higher than on Pd(111).
In the latter two cases, structural optimization leads to a remarkable 
reconstruction of the adsorbate layer (Fig.~\ref{CH4_config_Pd}(a)).
Both the CH$_3$ group and the H atom move away from their initial adsorption
sites and bind to the same O atom, thus forming an adsorbed methanol molecule.
Notably, if CH$_3$ and H are initially placed over other adsorption sites on Pd(111), 
CH$_3$ and H also detach from the metal atoms, but bind separately to different O atoms 
of the surface, resulting in adsorbed OH and CH$_3$O groups and no methanol formation.
The same is found in the case of the Pt(111) surface where CH$_3$ and H remain bound to
separate sites.
In this case, the Pt atoms to which CH$_3$ is bound are lifted
off the surface layer by 2.2 to 2.5 \AA, depending on the initial configuration
(cf. Fig.~\ref{CH4_config_Pt}).

Analogous calculations of methane adsorption energies have been performed for surfaces covered with 0.75~ML oxygen on the surface and 0.25~ML in sub-surface positions.
Again, three different initial positions of adsorbed CH$_3$ and H have been chosen: 
H always on top of oxygen, and 
(i) CH$_3$ on top of metal atom, 
(ii) CH$_3$ on top of oxygen atom, and 
(iii) CH$_3$ on fcc hollow site (oxygen vacancy). 
The largest adsorption energy results for CH$_3$ and H initially on top of oxygen 
for all considered surfaces (denoted by CH$_3$-fcc \& H-fcc in Table~\ref{diss}).
The energy values in Table~\ref{diss} indicate that methane adsorption on surfaces 
with sub-surface oxygen is  about 1~eV weaker than for the case of oxygen adsorbed 
solely {\it on} the surface.
In the presence of sub-surface oxygen, methanol does not form spontaneously starting from
the initial geometries above, H and CH$_3$ remaining separately bound to different oxygen atoms.

While in the simulations so far the formation of methanol takes place only on 
Pd(111) and Pd/Pt/Pd(111) with oxygen on-surface coverage, we now calculate the 
adsorption energy after methanol formation on all other surfaces, irrespective of 
the initial geometry choosen. 
To this end, a novel set of DFT structural relaxations is performed starting with methanol
adsorbed on all surfaces, including those with subsurface oxygen, using the atomic configuration obtained for Pd(111).
In all cases, the adsorption energies are higher than for separately adsorbed
CH$_3$ and H, as reported in Table~\ref{diss}.
As visible in Fig.~\ref{CH4_config_Pd}(a), the methanol molecule remains loosely 
adsorbed to the surface, mainly via a hydrogen-bond between the OH group of methanol 
and a surface O atom.
The methanol adsorption energies amount to -0.11 eV on Pd(111) and Pd/Pt/Pd(111), 
and -0.08 eV on Pt(111) for 1 ML on-surface oxygen coverage, 
reflecting weak molecule-surface interactions.

\subsection{CH$_4$ adsorption on superficial oxide layers}

\begin{figure}[ht!]  
\begin{center}
\includegraphics[clip,width=9cm,height=10.5cm]{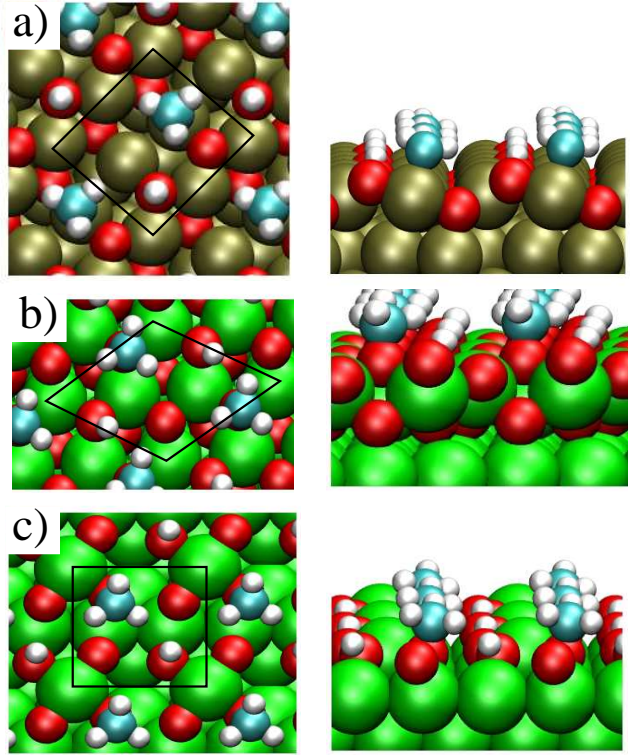}
\end{center}
\caption{
Optimized atomic structures of adsorbed CH$_3$ and H on thin oxide layers of
pure a) PdO(101)/Pd(100), b) $\alpha$-PtO$_2$(0001)/Pt(111),  and c) Pt$_3$O$_4$(100)/Pt(100). 
The shown atomic configurations correspond to the highest adsorption energy for each oxide layer.
Distances between adsorbed atoms and nearest neighbor substrate atoms in~\AA: 
a) d$_{C-Pd}$ = 2.05, d$_{C-O}$ = 2.77, d$_{H-O}$ = 0.98; 
b) d$_{C-Pd}$ = 3.15, d$_{C-O}$ = 1.43, d$_{H-O}$ = 0.98; 
c) d$_{C-Pd}$ = 2.85, d$_{H-O}$ = 0.98, d$_{C-O}$ = 1.44.
}
\label{CH4_config_Thin}
\end{figure}

\begin{figure}[ht!]
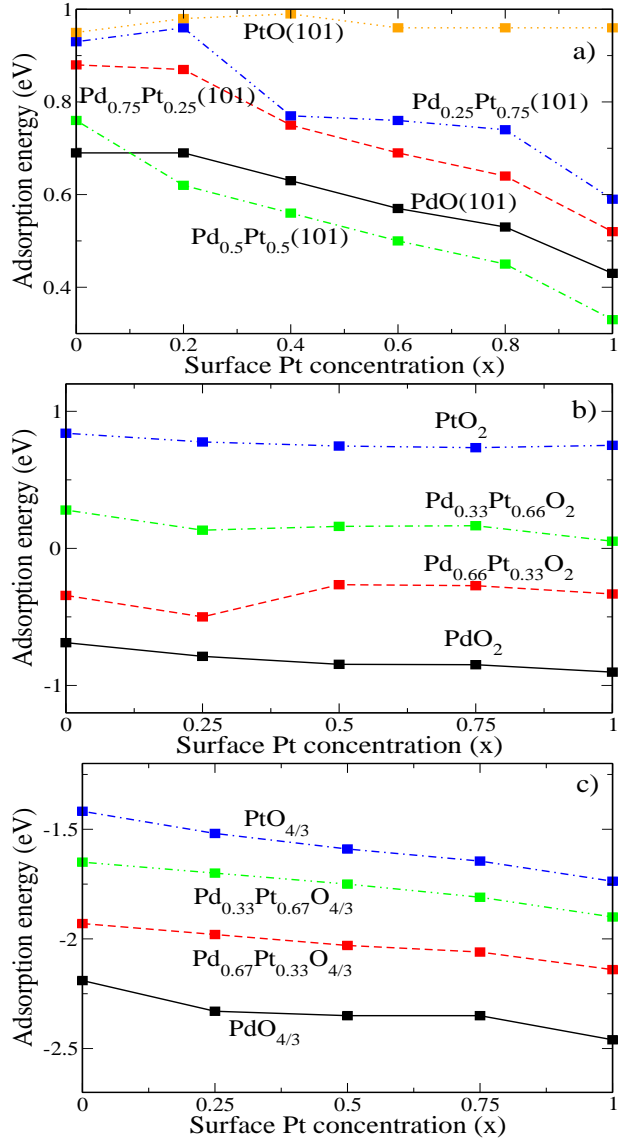
  
\begin{center}
\includegraphics[clip,width=8cm,height=5cm]{Fig6a.eps}\\
\includegraphics[clip,width=8cm,height=5cm]{Fig6b.eps}\\
\includegraphics[clip,width=8cm,height=5cm]{Fig6c.eps}
\end{center}
\caption{
Calculated methane adsorption energies on thin Pd-Pt oxide layers as a function 
of the Pt concentration $x$ in the metal surface layer with the Pt concentration $y$ 
in the oxide as a parameter:
(a) Pd$_{1-y}$Pt$_y$O(101)/Pd$_{1-x}$Pt$_x$/Pd(100), 
(b) $\alpha$-Pd$_{1-y}$Pt$_y$O$_2$/Pd$_{1-x}$Pt$_x$/Pt(111), and 
(c) Pd$_{1-y}$Pt$_y$O$_{4/3}$/Pd$_{1-x}$Pt$_x$/Pt(100).
}
\label{CH4_diss_Thin}
\end{figure}

With increasing oxygen supply, the oxidation of the metallic catalyst is expected
to proceed with formation of thin oxide layers. 
Correspondingly, we further investigate methane adsorption
on the thin oxide layer structures described in Sect. 3.2. 
The most favorable adsorption sites of methyl and hydrogen are determined 
via relaxation of different structures with the
following initial positions of the adsorbates:
both CH$_3$ and H on top of oxygen or metal atoms,
CH$_3$ on top of oxygen and H on top of metal atom, and vice versa.
The optimized atomic structures on the PdO(101)/Pd(100), 
$\alpha$-PtO$_2$(0001)/Pt(111), and 
Pt$_3$O$_4$(100)/Pt(100) layers are shown in Fig.~\ref{CH4_config_Thin}.
The Pd-Pt composition is varied both in the oxide layer and the outermost layer 
of the metal substrate (cf. Fig.~\ref{CH4_diss_Thin}).
For all compositions, the stable atomic configurations of adsorbed CH$_3$ and H
have been found to be qualitatively equal. 
However, the values of the adsorption energy, presented in Fig.~\ref{CH4_diss_Thin}, 
show comparatively large variations. 
Every curve point in Fig. 6 represents a special Pd-Pt configuration in the 
oxide and metal surface layer as described above. 
Test calculations of another Pd-Pt configuration for a PdO(101)- and PtO$_2$-like 
oxide layer at fixed composition (see Sect. 3.2) yield only small differences in 
the methane adsorption energy of less than 0.05~eV.

For the PdO(101)-like oxide layers, the most stable adsorption sites for CH$_3$ are 
either on Pd, if only Pd atoms are present, or on Pt, if Pt replaces Pd atoms 
in the oxide (Fig.~\ref{CH4_config_Thin}(a)).
The corresponding adsorption energy values 
(Fig.~\ref{CH4_diss_Thin}(a)) indicate that the methane decomposition reaction 
on this oxide structure is endothermic for all Pd-Pt compositions considered.
As a general trend, for a given composition of the oxide layer, the reaction is the 
more endothermic the more Pd atoms are present in the outermost layer of the metal
substrate, except for pure PtO(101) where the adsorption energy is independent of the
substrate composition.
For a given substrate composition, increasing the Pd/Pt ratio in the oxide 
from 0 to 1 results in a rather complex behavior of the adsorption energy 
values, whereby pure PtO(101) layer is the least reactive structure in all 
cases.

In the case of the $\alpha$-PtO$_2$-like layer, where all metal atoms are fully
coordinated by oxygen, adsorption of both CH$_3$ and H occurs necessarily on
the O atoms (Fig.~\ref{CH4_config_Thin}(b)).
The adsorption energy depends only weakly on the Pt concentration in the outermost
substrate-layer, whereas a strong dependence on the oxide layer composition is observed
(Fig.~\ref{CH4_diss_Thin}(b)).
While the reaction is clearly endothermic for the pure PtO$_2$ oxide layer,
it becomes exothermic for a Pd/Pt ratio in the oxide larger than about 0.5.
Thus, the driving force for methane dissociation increases monotonously with the Pd
content in the oxide.

In the case of the Pt$_3$O$_4$-like oxide layers, both CH$_3$ and H adsorb 
strongly on the undercoordinated exposed oxygen atoms (Fig.~\ref{CH4_config_Thin}(c)),
as found also previously for a pure Pt$_3$O$_4$ layer~\cite{Nicola}.
The adsorption energy of -1.74~eV obtained here is comparable to the value 
of -1.47~eV computed in Ref.~\cite{Nicola} using norm-conserving pseudopotentials.
Increasing the Pd concentration in the oxide layer leads to stronger and stronger
adsorption, with little dependence on the composition of the outermost substrate-layer 
(Fig.~\ref{CH4_diss_Thin}(c)).
In particular, the large adsorption energy of -2.5~eV, computed for 
the Pd$_3$O$_4$/Pt(100) oxide, 
is comparable to the driving force for methane dissociation and methanol
formation computed for the case of an oxygen ML adsorbed on the Pd/Pt/Pd(111) surface
(see Table~2).

\subsection{CH$_4$ adsorption on bulk PdO-like oxides}

After considering methane adsorption on thin oxide layers on Pd-Pt metal substrates, 
it is interesting to compute, for comparison, the methane adsorption energies
on the corresponding bulk oxide surfaces.
We note that for $\alpha$-PtO$_2$ the adsorption on a single layer is 
already representative of the behavior of the bulk oxide, which consists of 
stacked PtO$_2$ layers weakly bound by van der Waals forces~\cite{Nicola}.
Furthermore, the surface of bulk Pt$_3$O$_4$ with lowest surface energy is 
nearly identical to the surface of the thin oxide layer considered above.
In the following, we thus consider only surfaces of mixed Pd-Pt oxides with the 
structure of the well-known PdO bulk oxide phase.
In particular, methane adsorption energies are calculated on the (100) and (101)
surfaces, which present low surface energies~\cite{Rogal}. 

\begin{figure}[ht!]  
\begin{center}
\includegraphics[clip,width=6cm,height=8cm]{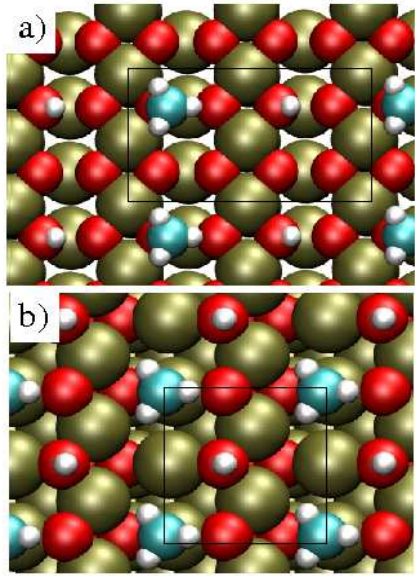}
\end{center}
\caption{
Optimized atomic structures of adsorbed CH$_3$ and H on the a) (100) and 
b) (101) surface of the bulk oxide phase PdO. 
Distances between adsorbed atoms and nearest neighbor substrate atoms in~\AA:
a) d$_{C-Pd}$ = 2.96, d$_{C-O}$ = 1.47, d$_{H-O}$ = 0.99; 
b) d$_{C-Pd}$ = 2.04, d$_{C-O}$ = 2.81, d$_{H-O}$ = 0.98.
}
\label{Thin_diss_str}
\end{figure}

As a model of a mixed oxide Pd-Pt-O, we analyze a slab of five PdO layers
covered with a mixed Pd$_{1-x}$Pt$_x$O layer, in which Pd and 
Pt are arranged in a chess-like pattern for $x=0.5$
(adsorption energies for the row-like Pd-Pt pattern differ less 
than 0.025~eV from the chess-like one).
Our analysis shows that on the (100) surface CH$_3$ and H adsorb preferentially 
on top of oxygen atoms, in agreement with previous calculations \cite{blancorey} 
(Fig.~\ref{Thin_diss_str}(a)).
On the (101) surface, H adsorbs on top of oxygen and CH$_3$ on top of metal atoms,
preferentially on Pt in the case of a mixed Pd-Pt bulk oxide, as also found for 
the thin oxide layer in the previous section (Fig.~\ref{Thin_diss_str}(b)).
The calculated adsorption energies on the different surfaces are listed 
in Table~\ref{diss_PdO}.
With increasing Pt content in the oxide, a monotonously decreasing driving force for 
methane adsorption is found on PdO(100), whereas on Pd(101) the adsorption energy 
displays a non-trivial dependence on the Pt content, similarly as in the case of the 
thin Pd(101)-like oxide layer considered above.
Unlike the case of thin PdO-like oxide layers, where methane adsorption is endothermic,
for the PdO bulk phase, the adsorption reaction is exothermic.
Regarding the dependence of the adsorption energy on the oxide composition, 
the largest value of -1.0~eV is obtained for the (100) surface of pure PdO
(cf. Fig.~\ref{CH4_config_Thin}(a)).
This value is however considerably lower than the energy values calculated for the 
Pt$_3$O$_4$-like oxide layers on Pt(100) and for the oxygen adlayers on Pd(111) 
(cf. Fig.~\ref{CH4_diss_Thin}(c) and Table~2).

\section{Discussion}

The dissociative adsorption of methane on noble metal surfaces represents an 
essential reaction step of the catalytic combustion of methane, and is thought 
to limit the reaction kinetics.
As a first effort towards a deeper understanding of this catalytic reaction,
we have studied methane adsorption on various oxide structures of Pd, Pt 
and Pd-Pt alloy surfaces. 
The choice of the investigated systems is motivated by the facts that palladium 
oxide phases, in particular PdO, have been suggested to be catalytically more 
active than pure Pd~\cite{Carlsson,Burch1}, and that mixed Pd-Pt catalysts 
have been found to possess higher conversion efficiency for methane 
combustion, especially on the long term.
Our investigation includes chemisorbed O atoms, thin oxide layers and bulk 
PdO-like oxide surfaces, addressing both their thermodynamic stability and 
their reactivity towards methane dissociation.

\subsection{Formation of oxide layers on Pd-Pt surfaces}

The thermodynamic stability of different oxidized Pd-Pt structures has been
addressed by calculating average oxygen binding energies at 0~K. 
In general, superficial oxide layers are found to be more stable than oxygen 
adlayers(including sub-surface O atoms) at the corresponding oxygen coverages, 
as displayed in Table~2 and Fig.~\ref{Binding_Oxide}.
The larger stability of oxygen adlayers on Pd compared to those on Pt can be  
understood simply from the lower electronegativity of palladium.
Consistently with previous findings~\cite{Dianat1}, the largest binding energy 
is found for a Pd/Pt/Pd(111) layer stack, due to the partial donation of electrons
from the Pd surface layer to the Pt sub-surface layer, which increases the
surface reactivity.

In the case of surface oxide layers, the Pd-Pt composition of the outermost 
layer of the metal substrate has only a minor influence on the oxygen binding 
strength.
Changes of the binding strength due to varying the Pd/Pt ratio 
in the oxide can be explained on the basis of the stability of the corresponding bulk
oxide phases, as thoroughly addressed in Refs.~\cite{Nicola,Dianat2}.
Namely, mixed oxides with PdO structure are destabilized by replacing
Pd atoms with Pt, while mixed oxides with $\alpha$-PtO$_2$ or Pt$_3$O$_4$
structure are destabilized by replacing Pt atoms with Pd.

\subsection{Dissociative adsorption of CH$_4$ on oxidized Pd-Pt surfaces}

Concerning the dissociative adsorption of methane, our DFT calculations at 
the GGA level suggest, in agreement with existing literature~\cite{Paul}, that 
Pd metal surfaces are not reactive, and Pt surfaces are only little reactive. 
Since CH$_3$ and H possess an electron-donor character, 
they are expected to bind better to electronegative elements.
Indeed, binding to pure Pt (electronegativity 2.3 on the Pauling scale)
is stronger than to pure Pd (electronegativity 2.2).
However, if adsorbed oxygen (electronegativity 3.4) is present on the
surface, either in the form of an oxygen adlayer or as surface oxide,
then binding to oxygen provides a strong driving force for methane 
dissociation.

Correspondingly, we have obtained the largest adsorption energies for 1~ML of
oxygen adsorbed on the (111) metal surfaces, in particular on Pd(111) and 
on the Pd/Pt/Pd(111) layer stack, with an energy gain of about -2.4 eV (cf. Table 2). 
A peculiar effect observed on the latter two surfaces 
is the spontaneous formation of a methanol molecule which remains only loosely bound 
to the surface via hydrogen bonds and weak metal-oxygen interactions 
(Fig.~\ref{CH4_config_Pd}(a)).
We observe a net thermodynamic preference for methanol formation on all
three surfaces considered, with the largest values being obtained on
Pd(111).
An analysis of the energy barriers associated with the
CH$_4$ $\rightarrow$ CH$_3$OH conversion reaction exceeds
the scope of the present investigation.
However, the possibility of a direct methane to methanol conversion on the 
(111) surface of transition metals has been recently put forward in a 
theoretical DFT study~\cite{fratesi}.
It has to be noted, however, that the further combustion of methanol on
Pd-Pt surfaces is expected to take place at the temperatures required
to dissociate the C-H bonds of methane.
Therefore, in general the selectivity towards methanol formation is found
to decrease by increasing the overall conversion efficiency~\cite{holmen}. 

In the case of superficial oxide layers formed on Pd-Pt, we have found that methane 
adsorption is clearly endothermic both on the thin PdO(101)-like layer on (100) surfaces 
and on the thin $\alpha$-PtO$_2$-like layer on (111) surfaces.
Since the same was noted for oxidation of carbon monoxide \cite{Gao}, these phases may 
be considered as a kind of passivation layer on the metal substrate, whose formation
may suddenly reduce the oxidation activity of the catalysts.
This property, however, is not shared by the surfaces of bulk oxides, where both the (101) and the (100) surfaces are reactive towards methane dissociation (Table 3).
This behavior can be understood by considering the charge transfer 
between the CH$_3$ molecule and the surface atoms.
According to a Bader analysis, in the case of bulk PdO, the CH$_3$ molecule donates 
electrons to the surface and becomes positively charged (+0.44e), while in the case 
of the PdO-like oxide layer CH$_3$ gains electrons (-0.13e).
This indicates a strongly reduced electronegativity of the metal atoms of the
thin oxide layer due to the presence of the underlying metal substrate, compared with
the surfaces of bulk oxides.
As a consequence, binding of CH$_3$ to the surface of the bulk oxide is favorable, 
in contrast to the oxide monolayer over the metal substrate.
This finding is consistent with experimental observations that oxidation of Pd with 
formation of PdO is beneficial for methane oxidation \cite{Carlsson,Burch1}. 
In comparison, bulk $\alpha$-PtO$_2$ is inert with respect to methane dissociation in the absence of defects (see also Ref.~\cite{Nicola}).
This again is consistent with observations that formation of bulk oxide phases 
is beneficial for methane oxidation on Pd, but not on Pt \cite{Carlsson, Becker, Burch1}.
Oxidation of Pt might be beneficial if the reaction conditions allow the
formation of Pt$_3$O$_4$-like phases (see also the discussion in 
\cite{Nicola, Nicola2, Nicola3}).
Namely, our DFT calculations predict adsorption energies of CH$_3$ and H
on Pt$_3$O$_4$ and Pd$_3$O$_4$ which are higher than those on PdO surfaces
and comparable with that on oxygen adlayers covering Pd(111).
This is due to the availability of undercoordinated oxygen sites on the exposed
Pt$_3$O$_4$(100) surface, to which electron acception from CH$_3$ and H
is very favorable.

Concerning effects on methane adsorption resulting from Pd-Pt alloying,
we note that in most of our calculations the composition
of the metal substrate underneath the thin oxide layer does not show
a pronounced effect on the methane adsorption energy.
However, the effect of the composition of the oxide layer is evident
and the reactivity decreases with increasing Pt content.
An exception is the thin PdO(101)-like oxide layer on (100) surfaces, where
the CH$_3$ group binds preferentially to the metal atom rather than to oxygen.
In this case, the presence of Pt atoms does increase the reactivity since
they are stronger electron acceptors, as mentioned above.
However, at the same time the presence of Pt results in a decrease of the
reactivity of the O atoms towards the adsorption of H.
These two counteracting effects result in a non-trivial trend of the computed 
adsorption energies with increasing Pt content, both in the case of
the thin layers (Fig.~\ref{CH4_diss_Thin}) and of bulk PdO(101) (Table~3).
The actual adsorption energies in this case depend on the specific
arrangements of Pt, Pd and O atoms close to the adsorption sites of CH$_3$
and H, both in the oxide layer and in the metal substrate.
However, in none of the cases considered is the adsorption reaction
exothermic, as discussed previously.

\section{Conclusions}

In conclusion, we have performed extensive calculations of the driving
force for methane dissociation on clean and oxidized Pd-Pt surfaces.
Both in the case of Pd and Pt, formation of stable thin layer oxide structures
such as the PdO(101)-like layer on the (100) surface and the $\alpha$-PtO$_2$-like 
layer on the (111) surface, leads to a reactivity loss towards the dissociative
adsorption of methane.
Methane adsorption is instead favored on metastable surface oxide structures
such as a Pt$_3$O$_4$-like layer, whose reactivity increases with increasing Pd
content.
Furthermore, reactivity is recovered for exposed surfaces of {\em bulk} PdO, 
consistently with existing experimental results~\cite{Lapisardi,Yammamoto}.

In the case of oxygen adlayers we have found that the formation 
of a methanol molecule after methane dissociation is thermodynamically 
favored.
This may suggest that selective conversion of methane to methanol  
rather than total oxidation can be achieved under pressure and  
temperature conditions that prevent the formation of surface oxides.
However, the further combustion of the produced methanol over other sites 
of the Pd-Pt surface cannot be excluded, reducing the selectivity of the
conversion reaction.

As a final remark, we would like to note that our investigation, being concerned with the thermodynamic driving force for the methane dissociation reaction, should be considered only as a preliminary indication about the reactivity of different surface structures.
Based on the results presented here, more thorough calculations of the corresponding activation barriers, which can be directly related to the kinetic constants for adsorption, shall be the subject of forthcoming works.

\subsection*{Acknowledgments}

Computational resources were provided by the Center for Information Services
and High Performance Computing (ZIH) of the Technische Universit\"at Dresden.
This work was partially supported by the Deutsche Forschungsgemeinschaft
under contract CI~144/1-2 and CU~44/9-2.
LCC acknowledges support by the Deutsche Forschungsgemeinschaft within the
Emmy Noether Programme.
NS acknowledges support by the FWF.

\newpage

\section*{Tables}

\begin{table}[h!]  
\caption{
Calculated average oxygen binding energies (in eV) in the case of only on-surface 
adsorption (on) and for simultaneous on-surface and sub-surface adsorption (on+sub) 
on Pd(111), Pt(111), and Pd/Pt/Pd(111) at different O coverages.
In the on+sub case, the amount of sub-surface oxygen is always 0.25 ML and the 
remaining oxygen is on the surface, i.e. for 0.25 ML total coverage there is no 
oxygen on the surface in the on+sub case. 
}
\begin{center}
\begin{tabular}{c|cc|cc|cc}
\hline
Total O coverage & \multicolumn{2}{c|}{Pd(111)} & \multicolumn{2}{c|}{Pt(111)} & \multicolumn{2}{c}{Pd/Pt/Pd(111)}\\
in ML &  on\hskip -2mm\  & on+sub\hskip -2mm \
&  on\hskip -2mm\  & on+sub\hskip -2mm \
&  on\hskip -2mm\  & on+sub\hskip -2mm \\
\hline
0.25 & -1.34 & 0.36  &-1.22  & 0.68 & -1.39& 0.64\\
0.50 & -1.03&-0.93 &-0.92 &-0.52 & -1.06& -0.85\\
0.75 & -0.61&-0.72 &-0.55 &-0.47 & -0.69& -0.68\\
1.00 &-0.22 &-0.42 &-0.17 &-0.27 & -0.25&-0.41\\
\hline
\end{tabular}
\end{center}
\label{on_sub_tab}
\end{table}

\newpage
\begin{table}[h!] 
\caption{
Calculated adsorption energies (in eV) for the dissociative adsorption of methane
on clean (111) and (100) surfaces of Pd, Pt, and Pd/Pt/Pd, and on surfaces covered 
with 1 ML oxygen {\it on} the surface, as well as with 0.75~ML on the surface 
and 0.25~ML in sub-surface positions. 
For clean surfaces, the values in squared brackets (in italics) correspond to
adsorption energies within LDA.
In the calculations, the initial positions of CH$_3$ and H have been 
chosen on high-symmetry sites of the (111) metal surface: 
CH$_3$ on top of metal atom (-top) and H on fcc hollow site on top of oxygen 
(-fcc) in case of the clean surfaces. 
For the systems with sub-surface oxygen, two fcc sites have been considered, 
CH$_3$ on fcc hollow site on oxygen atom (-fcc) and CH$_3$ on oxygen vacancy 
(-fcc*). 
The values in parentheses are the adsorption energies of methanol formation on 
the surfaces, irrespective of the reaction path. (These value are not related 
to the initial CH$_3$ and H positions in the table, and are arbitrarily reported 
besides the corresponding highest energy values obtained for the other
initial configurations).}
{\small
\begin{center}
\begin{tabular*}{\textwidth}{cccccccc}
\hline
    \multicolumn{8}{c}{Clean surfaces} \\
\hline 
Pd(111) & Pt(111) & Pd/Pt/Pd(111) & & & Pd(100) & Pt(100) & Pd/Pt/Pd(100) \\
\cline{1-4}
\cline{6-8} 
 0.19 {\it [-0.43]} & -0.08 {\it [-0.66]}  & 0.22 {\it [-0.45]} & & & 0.29  & -0.31  & 0.35  \\  
      \hline
      \hline
     \multicolumn{8}{c}{1~ML oxygen on (111) surfaces } \\
\hline
    &\multicolumn{2}{c}{CH$_3$-top \& H-fcc} & &  \multicolumn{2}{c}
{CH$_3$-fcc \& H-fcc} & \multicolumn{2}{c}{CH$_3$-hcp \& H-fcc} \\
    \cline{1-3}
      \cline{4-6}
       \cline{6-8}
      Pd(111) &\multicolumn{2}{c}{-2.42} & &  \multicolumn{2}{c}{-1.65} & 
\multicolumn{2}{c}{-1.39}\\ 
      Pt(111) &\multicolumn{2}{c}{{-2.13} (-2.27)} & &  \multicolumn{2}{c}
{-1.62} & \multicolumn{2}{c}{-1.99}\\
      Pd/Pt/Pd(111) &\multicolumn{2}{c}{-2.47} & &  \multicolumn{2}{c}{-1.74} 
& \multicolumn{2}{c}{-1.70}\\
\hline \hline
 \multicolumn{8}{c}{0.75~ML oxygen on-surface/0.25~ML oxygen sub-surface} \\
\hline
    &\multicolumn{2}{c}{CH$_3$-top \& H-fcc} & &  \multicolumn{2}{c}
{CH$_3$-fcc \& H-fcc} & \multicolumn{2}{c}{CH$_3$-fcc* \& H-fcc} \\
     \cline{1-3}
      \cline{4-6}
       \cline{6-8}
     Pd(111) &\multicolumn{2}{c}{-0.80} & &  \multicolumn{2}{c}{{-1.59} 
(-1.87)} & \multicolumn{2}{c}{-0.75}\\ 
      Pt(111) &\multicolumn{2}{c}{-1.39} & &  \multicolumn{2}{c}{{-1.40} 
(-1.78)} & \multicolumn{2}{c}{-1.23}\\
      Pd/Pt/Pd(111) &\multicolumn{2}{c}{-0.95} & &  \multicolumn{2}{c}{{-1.28} 
(-1.88)}& \multicolumn{2}{c}{-0.61}\\
\hline
\end{tabular*}
\end{center}
}
\label{diss}
\end{table}

\newpage  
\begin{table}[h!]  
\caption{
Calculated methane adsorption energies (in eV) on the (101) 
and (100) surfaces of the PdO
bulk oxide phase with one mixed oxide surface layer
Pd$_{1-x}$Pt$_{x}$O.
}
\begin{center}
\begin{tabular}{l|c|c|c|c|c}
\hline
Structure& x=0 & x=0.25 & x=0.5  & x=0.75 & x=1  \\
\hline
PdO(101) & -0.37 & -0.63 & -0.82 &-0.61 & -0.45 \\
PdO(100) & -1.00& -0.94 & -0.57& -0.51& -0.18\\
\hline
\end{tabular}
\end{center}
\label{diss_PdO}
\end{table}

\newpage

\section*{Figures}

\newpage

\newpage

\newpage

\newpage

\newpage

\newpage

\newpage

\end{document}